\documentclass{aa}  

\usepackage{graphicx}
\usepackage{lscape}
\usepackage{txfonts}
\begin{document}

   \title{HD\,344787: a true Polaris analogue? \thanks{Based on observations made with the Italian Telescopio Nazionale Galileo (TNG) operated by the Fundación Galileo Galilei (FGG) of the Istituto Nazionale di Astrofisica (INAF) at the Observatorio del Roque de los Muchachos (La Palma, Canary Islands, Spain).}}


\author{V. Ripepi \inst{1} 
          \and 
        G. Catanzaro \inst{2}
          \and 
        L. Moln\'ar \inst{3,4,5}
          \and
        E. Plachy \inst{3,4}
          \and
        M. Marconi \inst{1}
          \and
          G. Clementini \inst{6}
          \and
          R. Molinaro \inst{1}
          \and \\
          G. De Somma \inst{1,7,8}
          \and
          S. Leccia \inst{1}
          \and
        S. Mancino \inst{9,10,11}
        \and
         I. Musella \inst{1}
          \and
        F. Cusano  \inst{6}
            \and
         V. Testa \inst{12}
}

\institute{ INAF-Osservatorio Astronomico di Capodimonte, Salita Moiariello 16, 80131, Naples, Italy\\  \email{vincenzo.ripepi@inaf.it}
\and 
INAF-Osservatorio Astrofisico di Catania, Via S.Sofia 78, 95123, Catania, Italy \\
             \email{giovanni.catanzaro@inaf.it}
             \and
Konkoly Observatory, ELKH Research Centre for Astronomy and Earth Sciences, Konkoly Thege 15-17, H-1121 Budapest, Hungary
            \and
MTA CSFK Lend\"ulet Near-Field Cosmology Research Group
            \and
ELTE E\"otv\"os Lor\'and University, Institute of Physics, 1117, P\'azm\'any P\'eter s\'et\'any 1/A, Budapest, Hungary
             \and
INAF-Osservatorio di Astrofisica e Scienza dello Spazio, Via Gobetti 93/3, I-40129 Bologna, Italy 
             \and
Dipartimento di Fisica "E. Pancini", Università di Napoli "Federico II", Via Cinthia, 80126 Napoli, Italy
\and
Istituto Nazionale di Fisica Nucleare (INFN)-Sez. di Napoli, Via Cinthia, 80126 Napoli, Italy     
\and        
European Southern Observatory, Karl-Schwarzschild-Strasse 2, D-85748 Garching bei Munchen, Germany
\and
Cluster of Excellence Universe, Technical University of Munich, Boltzmannstr 2, D-85748 Garching, Germany
\and 
Ludwig-Maximilians-Universitat M\"{u}nchen, Physics Department, Geschwister-Scholl-Platz 1, D-80539 M\"{u}nchen, Germany
\and
             INAF – Osservatorio Astronomico di Roma, via Frascati 33, 
             }

   \date{}

 
  \abstract
   {Classical Cepheids (DCEPs) are the most important primary indicators for the extragalactic distance scale, but they are also important objects in their own right, allowing us to place constraints on the physics of intermediate-mass stars and the pulsation theories. }   
   {We have investigated the peculiar DCEP HD 344787, which is known to exhibit the fastest positive
   period change 
   of DCEPs, along with  a quenching amplitude of the light variation.}  
   {We used high-resolution spectra obtained with HARPS-N at the TNG for  HD 344787 and the more famous Polaris DCEP  to infer their detailed chemical abundances. Results  from the analysis of new time-series photometry of HD 344787 obtained by the TESS satellite are also reported.}
   {The double-mode nature of the HD344787 pulsation is confirmed by an analysis of the TESS light curve, although with rather tiny amplitudes of a few dozen millimag. This is indication that HD344787 is on the verge of quenching the  pulsation. Analysis of the spectra collected with HARPS-N at the TNG reveals an almost solar abundance and no depletion of carbon and oxygen. This means that the star appears to have not gone through  first dredge-up. Similar results are obtained for Polaris.    
   }
   {Polaris and HD344787 are both confirmed to be most likely at their first crossing of the instability strip. The two stars are likely at the opposite borders of the instability strip for first-overtone DCEPs with metal abundance  Z=0.008. A comparison with other  DCEPs that are also thought to be at their first crossing allows us to speculate that the differences we see in the Hertzsprung-Russell diagram might be  due to differences in the properties of the
   DCEP progenitors during the main-sequence phase.}


   \maketitle
%

\section{Introduction}\label{sect:intro}

Although they are primarily 
 known for their fundamental role in the extra-galactic distance scale, Classical Cepheids (DCEPs) also allow us to obtain insights into evolutionary properties and stellar interiors through their pulsational properties and thus to place constraints on the physics of intermediate-mass stars and pulsation theories \citep[see e.g.][and references therein]{Anderson2016,Bhardwaj2018,Marconi2020}.
In this context, DCEPs showing rapidly increasing periods  are particularly interesting because they are thought to be crossing the instability strip (IS) for the first time \citep[see e.g.][]{Turner2006}. Polaris is the most famous of these DCEPs. The star exhibits a fast-increasing period with changes of about 4.4–4.9 s yr$^{-1}$ \citep[see e.g.][]{Evans2002,Turner2005,Bruntt2008} along with abrupt variations  
\citep[see][]{Turner2009}. Polaris also has a peculiarly low and changing amplitude of the light variation, which  decreased from $\sim$0.1 mag during most of the past century to a few hundredths of magnitude in the early 2000s \citep[e.g.][]{Turner2009}. This amplitude is more typical of pulsators at the hot and cool edges of the strip, whereas Polaris was thought to be in the middle of the IS.  According to \citet{Turner2013}, Polaris observations can be explained if the stars is at its first crossing, taking into account that theoretical models  by \citet{Alibert1999} predicted a first-crossing IS shifted blueward, and in general, the DCEP IS tends to widen and becomes redder with increasing luminosity \citep[see e.g.][]{BCM2000,Desomma2020}, allowing the convection to damp  pulsation at hotter temperatures than at the other crossings. 
The redward path of Polaris towards the cool edge of the first crossing IS suggested by its constantly increasing period would thus also be  consistent with its decreasing amplitude. However, Polaris has not ceased pulsating just yet. After its amplitude reached a minimum approximately in 1990, it has slowly been increasing again \citep[e.g.][see section~\ref{sect:comparison} for more details about this point]{Bruntt2008,Turner2009}, although current radial velocity amplitudes have still not reached the levels measured in the first half of the twntieth century \citep{Usenko2018}.

A further object with Polaris-like characteristics was discovered 
a decade ago by \citet{Turner2010}: HD 344787. The star  
is a multi-mode DCEP, with  fundamental (F) and first-overtone (FO) periods of 5.4 days and 3.8 days, respectively. Study of the O--C diagram revealed that the F mode period of HD\,344787 is increasing by 
12.96$\pm$2.41 s yr$^{-1}$, which is one of the fastest period variations ever measured for a DCEP. According to \citet{Turner2010}, this period variation is in agreement with the expectation of stellar evolution theory for a first-crossing, redward-evolving DCEP. Even more striking is the similarity with Polaris for what concerns the light amplitude.  HD 344787 shows a quickly diminishing amplitude that decreased from $\sim$0.05 mag (summing F and FO amplitudes) in the first 20 years of the past century to become barely detectable approximately in 2008-2009.  
The waning amplitude would imply that the star is leaving the IS, and HD 344787 is foreseen to soon  completely cease pulsation, as the most recent observations collected by \citet{Turner2010} in 2010 found amplitudes compatible with non-variability ($\sim$2--3 mmag) from ground-based observations.

Notwithstanding the close similarity to Polaris and its clear relevance in the context of stellar evolution and pulsation theories, the study of HD\,344787 has not progressed after the work by \citet{Turner2010}. In this paper we 
continue the investigation of HD\,344787 based on three new 
pieces of information: 1) high-precision photometry of HD\,344787 by the TESS\footnote{Transiting Exoplanet Survey Satellite} satellite \citep{Ricker2015}, 2) elemental abundances from high-resolution spectra  obtained with the HARPS-N instrument at the TNG,  and  3) precise distances from the Gaia mission  \citep{Gaia2018}. 
The combination of these novel elements allows us to study the pulsation properties of HD\,344787 in detail and in turn to precisely locate the star in the Hertzsprung-Russell diagram.  
This allows us to discuss the evolutionary status of HD\,344787 in comparison with Polaris and other DCEPs that are thought to be in ther first crossings of the IS to explain the peculiarities shown by these stars.

Even though Polaris is the closest and brightest DCEP known so far, only a few chemical analyses of its atmosphere are available in the literature. In particular, the latest parameter determinations (effective temperature, gravity, and abundances) date back to~\citet{usenko05}. In order to obtain an updated set of abundances, 
we retrieved high-resolution spectra available for  Polaris in  the HARPS-N at the TNG archive, and to facilitate the comparison with HD\,344787, we 
analysed the spectra of both stars 
using exactly the same method (see Sect.~\ref{sect:dataAnalysis}).

The paper is organised as follows: Sect.~\ref{sect:observations} describes the spectroscopic observations and the data  reductions. TESS photometry for HD\,344787 is presented in Sect.~\ref{sect:tess}. In Sect.~\ref{sect:comparison} we discuss the comparison of H\,344787 with Polaris.
Finally, Sect.~\ref{sect:discussion} presents a discussion of our results, and Sect.~\ref{sect:conclusions} summarises the paper conclusions.

\section{Spectroscopic observations}\label{sect:observations}

\subsection{Observations and data reduction}
Three spectroscopic observations of the Galactic DCEP HD\,344787 were obtained at the 3.5m Telescopio Nazionale Galileo (TNG) equipped with the HARPS-N instrument in the nights of June 28 and July 15 and 22, 2020. 
The spectra cover the wavelength range between 3830 to 6930 {\AA}, with a spectral resolution R=115,000, and a signal-to-noise ratio (S/N) of about 100 at ${\lambda}$\,6000~{\AA} for each of them. 

Reduction of all spectra, which included  bias subtraction, spectrum extraction, flat fielding, and wavelength calibration,  was performed using the HARPS reduction pipeline. 
Radial velocities were  measured by cross-correlating each spectrum with a synthetic template. The cross-correlation was performed using the IRAF task {\it FXCOR} and excluding Balmer lines and  wavelength ranges, including telluric lines. The IRAF package RVCORRECT was used to determine heliocentric velocities by  correcting the spectra for the Earth’s motion.

We retrieved two spectra of Polaris from the HARPS-N archive. These spectra had been acquired consecutively on April 22,$^{}$ 2015. Both were taken with an exposure time of 20\,s over the wavelength range between 3830 to 6930 {\AA}, with a spectral resolution R=115,000, and have a signal-to-noise ratio (S/N) of about 250 at $\lambda$\,6000~{\AA}. Furthermore, we combined them and obtained an S/N higher than 300 over a spectral range greater than $\lambda$\,4000~{\AA}.

\subsection{Data analysis}
\label{sect:dataAnalysis}
Abundance analyses of HD\,344787 and Polaris followed exactly the same procedure. Effective temperatures were estimated using the line depth ratio (LDR) method \citep{Kovtyukh2000}.  LDRs have the advantage of being sensitive to temperature variations, but not to abundances and interstellar reddening. Typically, we measured about 32 LDRs in each spectrum.

\begin{table}
\centering
\caption{Atmospheric parameters. For each spectrum of HD\,344787 we list heliocentric Julian Day (HJD) at medium exposure (column 1), effective temperature (column 2), gravity (column 3), and the micro-turbulent and radial velocities (columns 4 and 5).}
\label{table_sum_spectro}
\begin{tabular}{ccccccc}
\hline \hline
      HJD  & T$_{\rm eff}$ & $\log g$ &    $\xi$      & v$_{\rm rad}$      \\
 2459000+  & (K)           &    (dex) & (km s$^{-1}$) & (km s$^{-1}$)      \\
 \hline                                                                             29.51441 & 5750 $\pm$ 150 & 1.3 $\pm$ 0.2 & 3.0 $\pm$ 0.2 & -8.67 $\pm$ 0.08\\
 46.46831 & 5750 $\pm$ 110 & 1.3 $\pm$ 0.2 & 3.1 $\pm$ 0.2 & -8.82 $\pm$ 0.09 \\
 53.71637 & 5750 $\pm$ 110 & 1.3 $\pm$ 0.2 & 3.0 $\pm$ 0.2 & -8.75 $\pm$ 0.06 \\
\hline                    
\end{tabular}
\end{table}

To determine the micro-turbulent velocity $\xi$, iron abundance, and $\log g$, we followed the iterative procedure recently outlined in \citet{catanzaro20}. Briefly, micro-turbulence was deducted  by the slope of iron abundances versus equivalent widths (EWs), while surface gravity was determined by imposing the ionisation balance between \ion{Fe}{I} and \ion{Fe}{II} lines \citep[145 \ion{Fe}{I} and 24 \ion{Fe}{II} spectral lines were extracted from][]{Romaniello2008}. EWs were measured using  an {\it IDL}\footnote{IDL (Interactive Data Language) is a registered trademark of Harris Geospatial Solutions} semi-automatic custom routine, which allowed us to minimise errors in the continuum evaluation on the wings of the spectral lines. Then they were  converted into abundances using the WIDTH9 code \citep{kur81} after generating an appropriate model atmosphere with the ATLAS9 LTE\footnote{local thermodynamic equilibrium} code \citep{kur93,kur93b}. The atmospheric parameters derived for each spectrum of HD\,344787 are summarised in Table~\ref{table_sum_spectro}. We obtained quite consistent parameters for the three spectra of HD\,344787, 
therefore we combined them in a  unique average spectrum with increased S/N$\approx$~130. For the subsequent abundance analysis we adopted the average values reported in the upper line of Tab.~\ref{param} for HD 344787. The lower line instead shows the parameter values  we obtained for Polaris.

\begin{table}
\centering
\caption{Atmospheric parameters adopted for HD\,344787 and Polaris.}
\label{param}
\begin{tabular}{lccc}
\hline \hline
Star  & T$_{\rm eff}$ & $\log g$ &    $\xi$     \\
      & (K)           &    (dex) & (km s$^{-1}$) \\
 \hline                                                                             HD\,344787 & 5750 $\pm$ 120 & 1.3 $\pm$ 0.2 & 3.0 $\pm$ 0.2 \\ Polaris    & 6000 $\pm$ ~70 & 2.0 $\pm$ 0.2 & 2.8 $\pm$ 0.2 \\
 
\hline                    
\end{tabular}
\end{table}

\begin{figure}
\centering
\includegraphics[width=8.5cm]{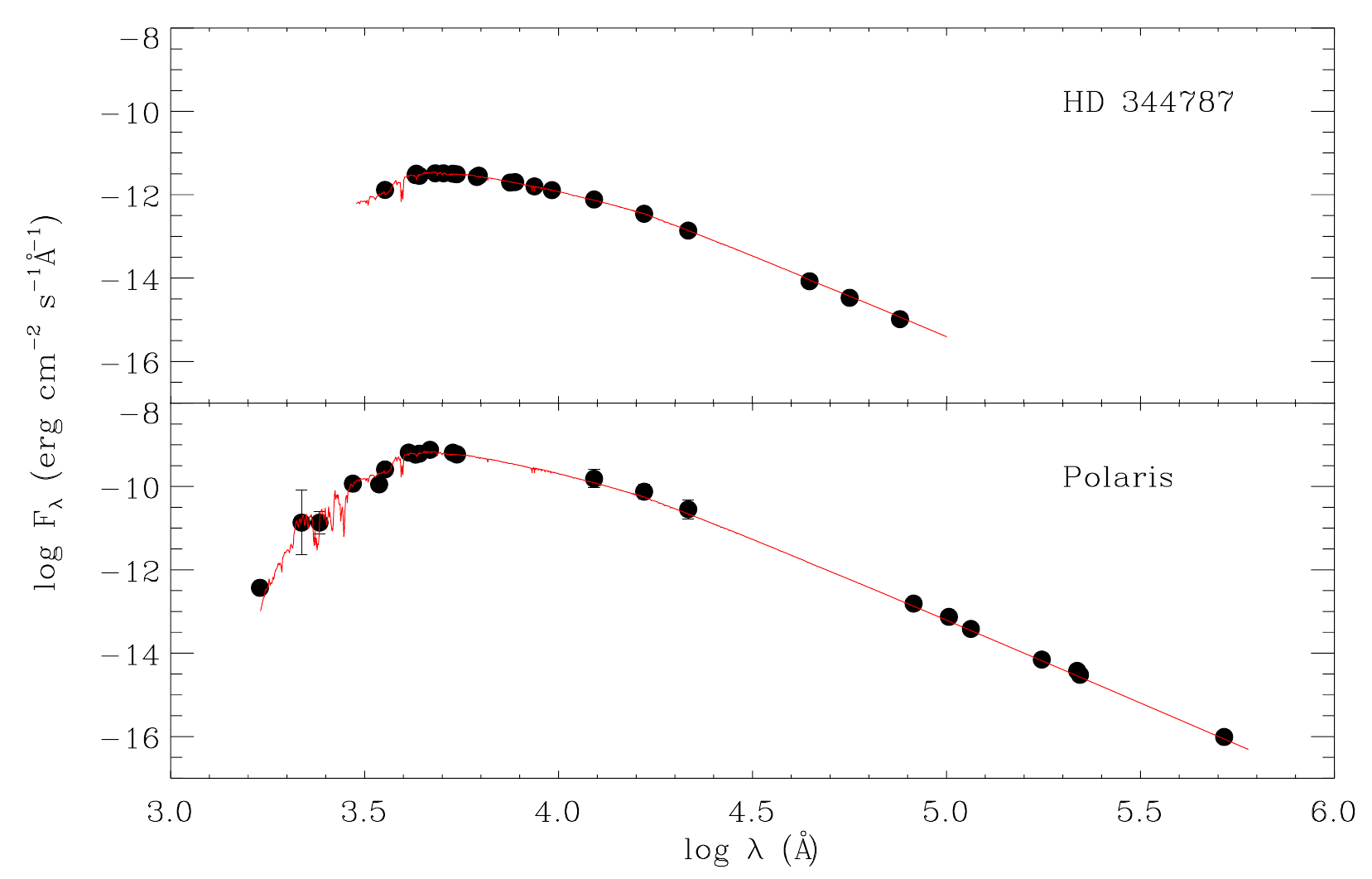}
\caption{Spectral energy distribution of HD\,344787 (upper panel) and Polaris (bottom panel). Filled dots represent the observed fluxes as retrieved from the VOSA tool. The red line shows  the theoretical flux computed using the ATLAS9 code.}
\label{sed}
\end{figure}

As a further check, we reproduced the observed spectral energy distributions (SEDs) with synthetic fluxes computed with the ATLAS9 code using the parameters reported in Table~\ref{param}. The observed fluxes were retrieved from the VOSA tool \citep{bayo08}, corrected for reddening by adopting $E(B-V)$\,=\,0.52\,$\pm$\,0.03 mag \citep[for HD\,344787]{Kervella2019} and $E(B-V)$\,=\,0.02\,$\pm$\,0.01 mag \citep[for Polaris]{Turner2013} and the \citet{fitz1999} extinction law. In Fig.~\ref{sed} we show the comparison between the observed and the theoretical SEDs. The upper panel refers to HD\,344787, where we excluded WISE photometry because it is contaminated by a close cool star, and the bottom panel shows Polaris.

Furthermore, using the distance inferred from the {\it Gaia} DR2 parallax \citep[$\varpi$\,=\,0.6865\,$\pm$\,0.0355 mas, to which we applied a zero-point correction of 0.049 mas;][]{Groenewegen2018}, we derived an absolute luminosity of  $L/L_\odot$\,=\,1435\,$\pm$\,147 for HD\,344787. For Polaris, we adopted the distance by Hipparcos (D\,=\,133\,$\pm$2 pc) and obtained $L/L_\odot$\,=\,2540\,$\pm$\,366. 

The radial velocity values listed in column 5  of Table~\ref{table_sum_spectro} were calculated by cross-correlating each observed spectrum with a synthetic template as described in the previous section. 
The 
three values are practically identical within the errors, and this could be interpreted as a vanishing amplitude  of the pulsation, thus  supporting a first crossing of the IS by HD\,344787. However, the almost identical radial velocities could also arise if the  three spectra had  been taken at  the same pulsation phase by chance.
 Using the pulsation modes (see Sect.~\ref{sect:tess}), we computed an artificial light curve of HD\,344787 covering the time interval  containing the Julian Dates of our spectroscopic  observations. This allowed us to verify that 
 the spectrum taken at HJD\,=\,2459046.46831 is close to a minimum of the HD\,344787 light curve,    whereas the spectrum  observed at HJD\,=\,2459053.71637 is close to a maximum. We can thus conclude that the almost null amplitude of the radial velocity variation of HD\,344787 is real.

\begin{figure*}
\includegraphics[width=16cm]{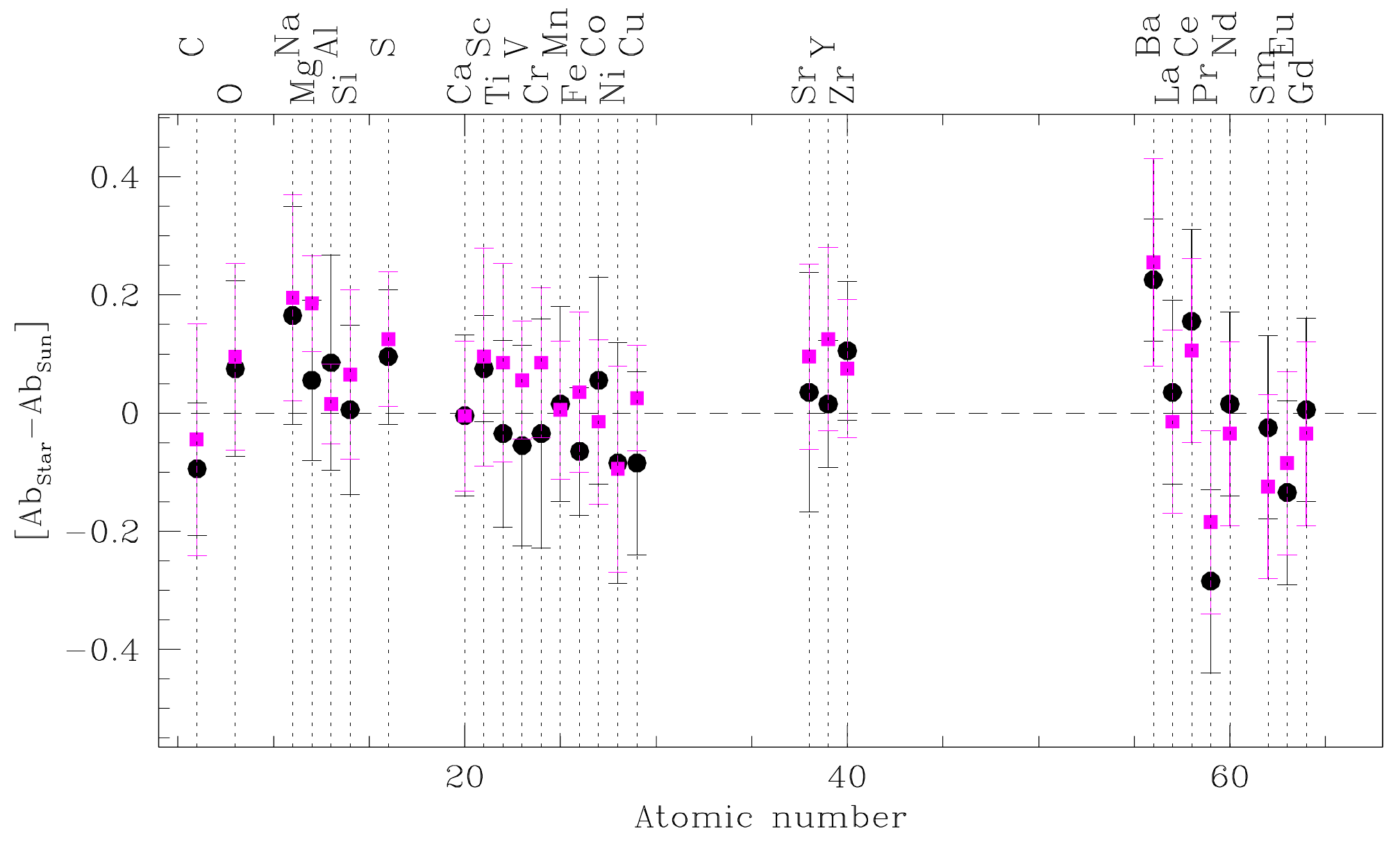}
\caption{Chemical pattern for HD\,344787 (black filled dots) and Polaris (magenta filled squares) with abundances plotted in terms of solar values  \citep{grevesse10}.}
\label{pattern}
\end{figure*}

Finally, the atmospheric parameters reported in Table~\ref{param} were used as input for the abundance analysis, which was performed following the procedures outlined in \citet{catanzaro19}. The total line broadening, estimated using metal lines, is 
12\,$\pm$\,1~km s$^{-1}$ for both stars. The abundances of the 28 species we detected in the spectra are provided in Table~\ref{abund} and are plotted with different symbols in Fig.~\ref{pattern}. The chemical patterns of HD\,344787 and Polaris are fully consistent within the errors.

Our elemental abundances  for Polaris can be compared 
with those derived by \citet{usenko05}. The only two differences concern magnesium and strontium. \citet{usenko05} found magnesium 
to be underabundant by 
$\approx$\,0.2~dex in comparison with the Sun, while we derived an overabundance of $\approx$\,0.2~dex, and  strontium 
is overabundant by about 0.4~dex for \citet{usenko05} while we found it to be  
consistent with the solar value. For all other species both analyses gave consistent results within the errors.

\begin{table}
 \centering
  \caption{Derived abundances for HD\,344787 and Polaris expressed in terms of solar values \citep{grevesse10}.}
  \begin{tabular}{lrr}
  \hline
  \hline
El  &      HD\,344787       &       Polaris ~~          \\
    &      [X/H]~~~~        &        [X/H]~~~~           \\
\hline
C   &  $-$0.09\,$\pm$\,0.11  &  $-$0.05\,$\pm$\,0.15 \\
O   &     0.07\,$\pm$\,0.14  &     0.09\,$\pm$\,0.20 \\
Na  &     0.16\,$\pm$\,0.18  &     0.18\,$\pm$\,0.16 \\
Mg  &     0.05\,$\pm$\,0.13  &     0.18\,$\pm$\,0.12 \\
Al  &     0.08\,$\pm$\,0.18  &     0.01\,$\pm$\,0.15 \\
Si  &     0.00\,$\pm$\,0.14  &     0.06\,$\pm$\,0.13 \\
S   &     0.09\,$\pm$\,0.11  &     0.12\,$\pm$\,0.11 \\
Ca  &     0.00\,$\pm$\,0.13  &     0.00\,$\pm$\,0.12 \\
Sc  &     0.07\,$\pm$\,0.08  &     0.09\,$\pm$\,0.15 \\
Ti  &  $-$0.03\,$\pm$\,0.15  &     0.08\,$\pm$\,0.17 \\
V   &  $-$0.05\,$\pm$\,0.17  &     0.05\,$\pm$\,0.10 \\
Cr  &  $-$0.03\,$\pm$\,0.19  &     0.08\,$\pm$\,0.15 \\
Mn  &     0.01\,$\pm$\,0.16  &     0.05\,$\pm$\,0.11 \\
Fe  &  $-$0.06\,$\pm$\,0.10  &     0.03\,$\pm$\,0.13 \\
Co  &     0.05\,$\pm$\,0.17  &     0.00\,$\pm$\,0.14 \\
Ni  &  $-$0.08\,$\pm$\,0.20  &  $-$0.09\,$\pm$\,0.17 \\
Cu  &  $-$0.08\,$\pm$\,0.15  &     0.02\,$\pm$\,0.09 \\
Sr  &     0.03\,$\pm$\,0.20  &     0.09\,$\pm$\,0.15 \\
Y   &     0.00\,$\pm$\,0.10  &     0.12\,$\pm$\,0.15 \\
Zr  &     0.10\,$\pm$\,0.11  &     0.07\,$\pm$\,0.11 \\
Ba  &     0.22\,$\pm$\,0.10  &     0.25\,$\pm$\,0.17 \\
La  &     0.03\,$\pm$\,0.15  &     0.00\,$\pm$\,0.15 \\
Ce  &     0.15\,$\pm$\,0.15  &     0.10\,$\pm$\,0.15 \\
Pr  &  $-$0.28\,$\pm$\,0.15  &  $-$0.18\,$\pm$\,0.15 \\
Nd  &     0.00\,$\pm$\,0.15  &  $-$0.03\,$\pm$\,0.15 \\
Sm  &  $-$0.00\,$\pm$\,0.15  &  $-$0.12\,$\pm$\,0.15 \\
Eu  &  $-$0.13\,$\pm$\,0.15  &  $-$0.08\,$\pm$\,0.15 \\
Gd  &     0.00\,$\pm$\,0.15  &  $-$0.03\,$\pm$\,0.15 \\ 
\hline\end{tabular}
\label{abund}
\end{table}

\section{TESS photometry}
\label{sect:tess}
HD\,344787 has recently been observed by the Transiting Exoplanet Survey Satellite (TESS). TESS was built to perform a nearly all-sky survey providing high-precision continuous photometric data that range from 27 days up to 351 days, depending on celestial position \citep{Ricker2015}. The southern ecliptic hemisphere was monitored in the first year of the mission, dividing the field of view into 13 overlapping sectors and rotating around the ecliptic pole. The telescope then turned to the northern ecliptic hemisphere in the second year and observed HD~344787 in Sector 14. The star has a brightness of 8.1~mag in the TESS pass-band, which spans the 600--1000 nm wavelength range and is centred on the Cousins \textit{I} band. At this brightness, a photometric precision per 30-minute cadence of  $\sim$100 ppm can be achieved with TESS, but instrumental issues strongly affect the data quality at certain observing times and positions within the field of view \citep{Huang2020}.

\subsection{TESS light-curve extraction}

\begin{figure}
\centering
\includegraphics[width=8.9cm]{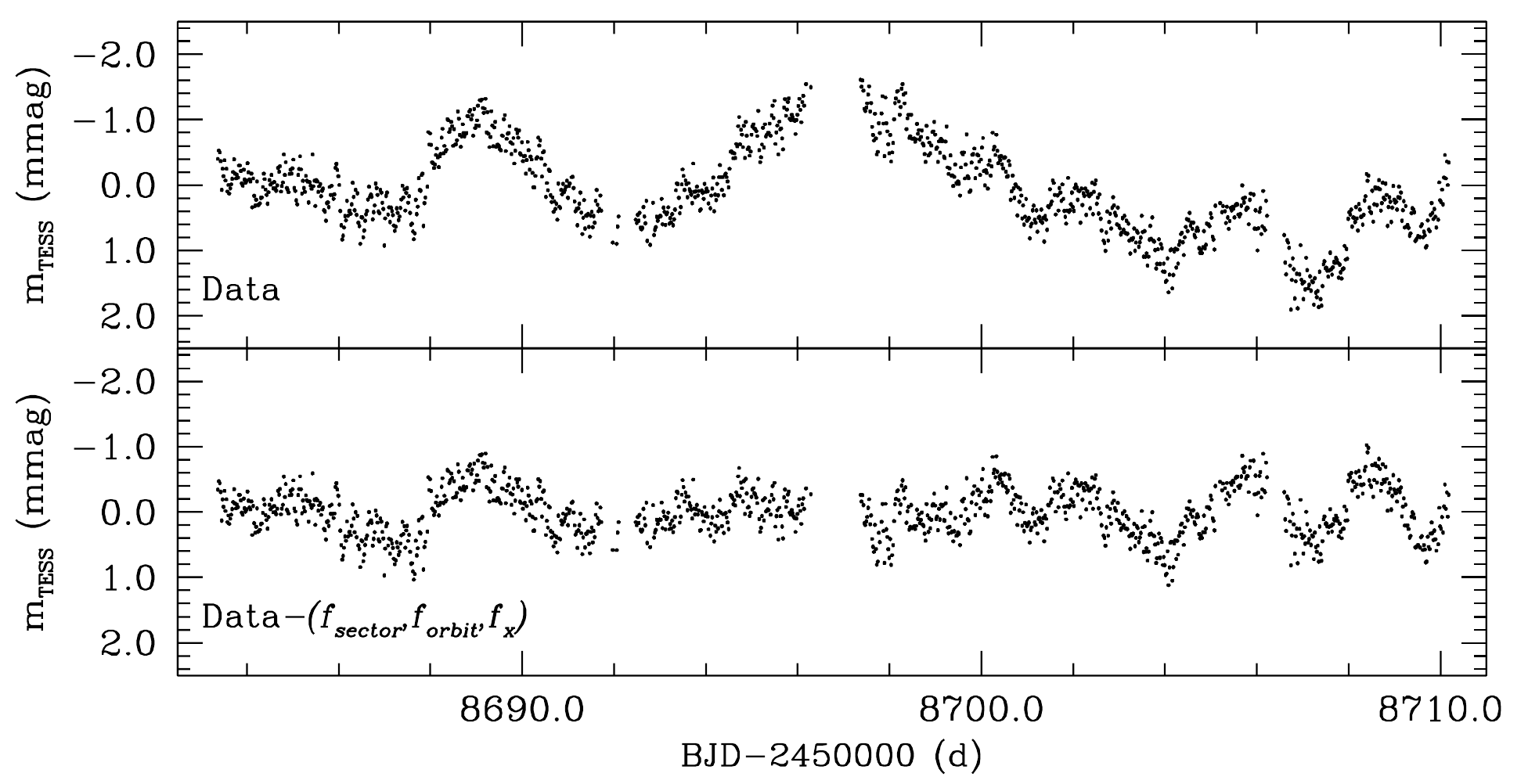}
\caption{TESS light curve of HD344787, as derived from the full-frame images with the FITSH differential image photometry pipeline. The top panel shows the light curve in mmag after subtracting the average value ($m_{TESS}$=8.1055 mag); the bottom panel shows the light curve after removing the first three instrumental frequencies (see Table~\ref{table_tess}).}
\label{fig:LC}
\end{figure}

HD 344787 was pre-selected to be observed in 2-minute cadence mode. These observations are made through small sub-images centred on the targets, and for these, photometric data are available at the Mikulski Archive, as provided by the TESS Science Processing Operations Center (SPOC). Unfortunately, no intrinsic variability could be detected in the SPOC light curve beyond a 20--30 mmag scatter, but it became evident that the star is affected by stray light and reflections from the metal straps of the electronics behind the CCD. Therefore we generated differential-image photometry with the FITSH package from the full-frame images that were taken with 30-minute cadence. The reduction process corrects many of the instrumental issues (differential velocity aberration, spacecraft jitters, background and stray light variations, strap reflection, etc.) and we found it to be extremely useful in previous studies of TESS variables \citep[see e.g.][]{Borkovits2020,Merc2020,Szegedi2020}. The pipeline also allows us to adjust the aperture size of the photometry, which is often critical in more crowded areas. TESS operates with a relatively low spatial resolution of 21“/px that results in significant contamination by nearby stars, especially in the denser stellar fields. Our previous tests indicated that an aperture with a 2.5 px radius works well for most of the stars, and we used this value for HD 344787 as well. Our photometric solution is displayed in Fig.~\ref{fig:LC}.
It is immediately clear from the light curve that any variation present has very small amplitude. 

\begin{figure}
\centering
\includegraphics[width=8.9cm]{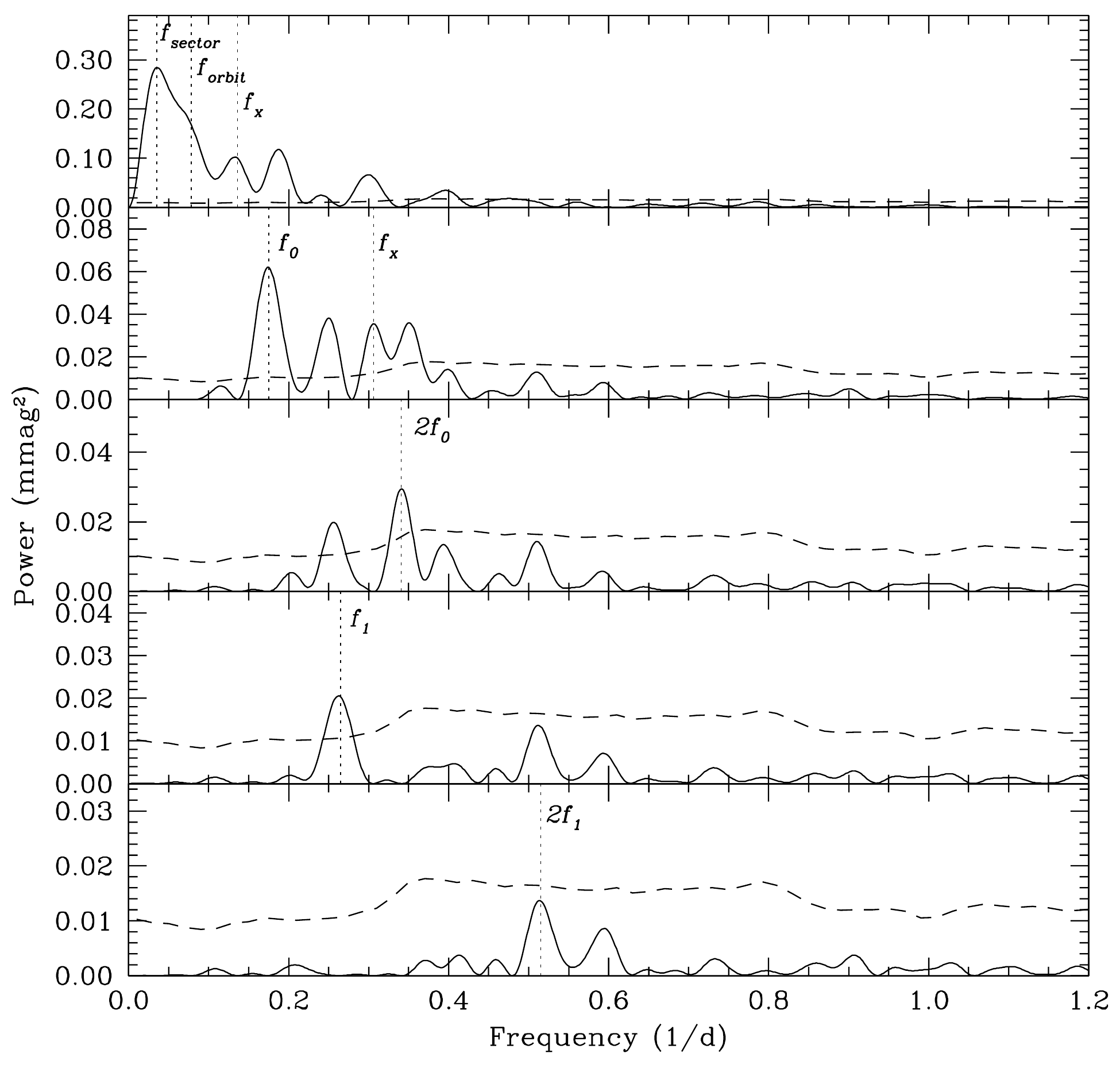}
\caption{Periodogram of HD344787 from TESS data. The different panels show the frequencies labeled in Table~\ref{table_tess}. The dashed line represents the noise level.}
\label{fig:Fou}
\end{figure}

\subsection{Frequency analysis}

\begin{figure}
\centering
\includegraphics[width=8.9cm]{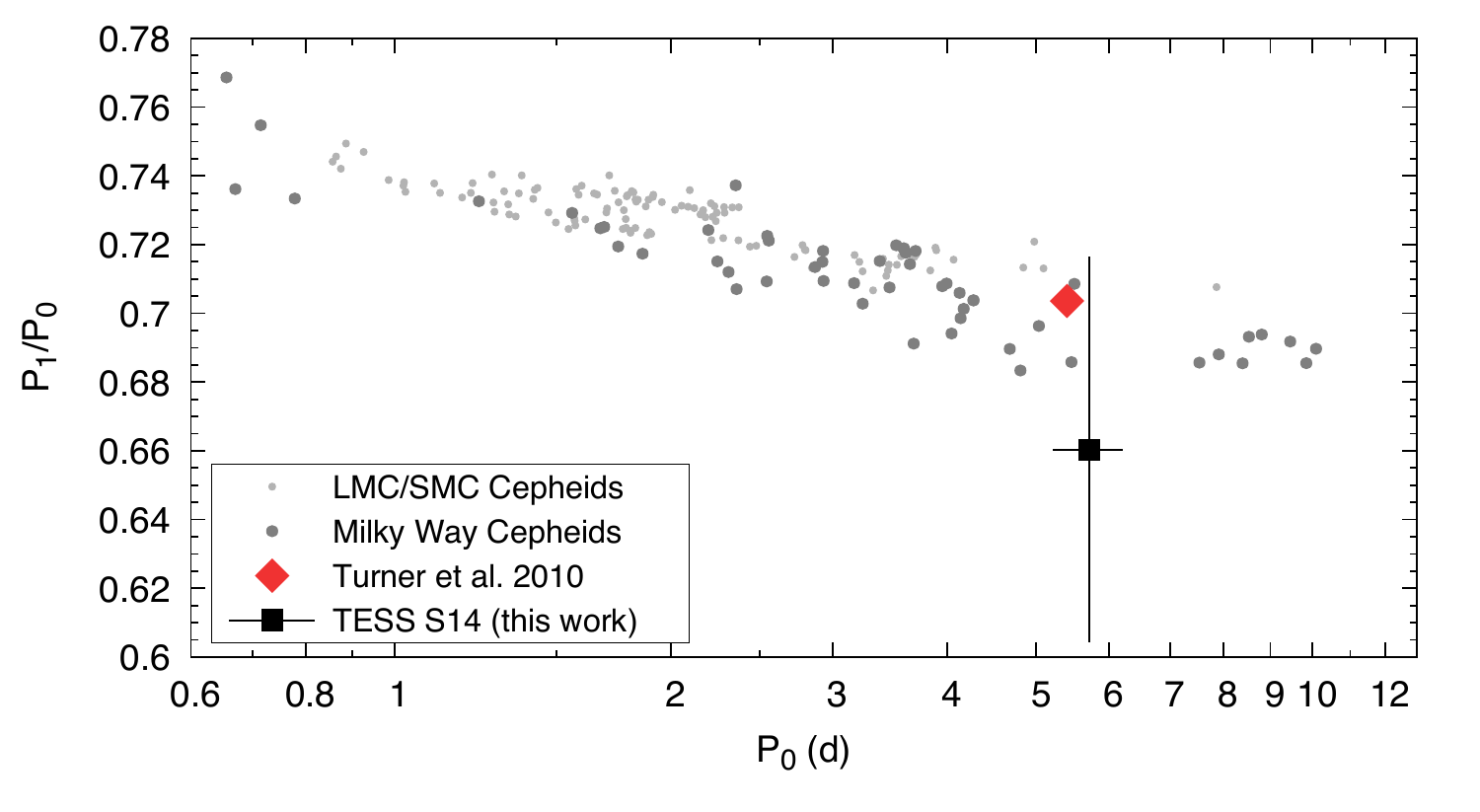}
\caption{Petersen diagram of double-mode DCEPs pulsating in F and FO modes, with the period ratios of HD344787 measured by \citet{Turner2010} (red diamond) and in this work (black square). Data for other Galactic DCEPs are from \citet{Udalski2018} and \citet{Jurcsik2018}. Data for DCEPs in the Large and Small Magellanic Clouds come from \citet{Soszynski2015}.}
\label{fig:Petersen}
\end{figure}

We performed a standard Fourier analysis with the Period04 software \citep{Lenz2005}  that revealed seven frequencies in the data with a signal-to-noise ratio (S/N) greater than 4. When we allowed a lower limit (S/N > 3), one more frequency can be detected. The frequency content is displayed in Table~\ref{table_tess} and graphically shown in Fig.~\ref{fig:Fou}. Several of the highest amplitude frequencies have instrumental origins: we identified the periodicity of the TESS orbit (13.7 days) and the length of a TESS sector (27.4 d, consisting of two orbits). The existence of these frequencies shows that our photometric pipeline did not entirely  eliminate instrumental issues, only reduced them to the mmag level. The double-mode pulsation causes a weaker but still detectable  signal in the TESS data: We found both the F mode ($f_0$) and the FO mode ($f_1$), according to the classification in the literature \citep[see][]{Turner2010}, along with the first harmonic frequencies (2$f_0$ and 2$f_1$). Both modes have shown decreasing amplitudes in the historical data of HD 344787 (Fig.~8 in \citealt{Turner2010}), but even with the uncertainties, variation in the amplitudes is noticeable and the F mode occasionally exceeds the FO mode that normally dominates in the light variation. This occurred again during the  TESS observations, in which we detected Fourier amplitudes that were twice as strong for the F mode than for the FO. We also found the additional frequency, $f_x$, and a possible combination frequency with $f_0$ ($f_0+f_x$). The notation x indicates that we are uncertain about the origin of this frequency, but given the low resolution of the TESS images, it is likely caused by blending. Interestingly, $f_x$  is close to the value of $0.5f_1$. 
Finally, we note that the period ratio of the main radial modes of HD 344787 is $P_0/P_1 \sim 0.70$ \citep[][]{Turner2009}. Considering that $P_0 \sim 5.4$ days, the position of the star in the Petersen diagram is in the expected region for Galactic DCEPs pulsating in both F and FO modes, as we show in Fig.~\ref{fig:Petersen}.

\begin{table}
\centering
\caption{Results of the  frequency analysis of HD 344787 based on TESS data obtained in Sector 14.}
\label{table_tess}
\begin{tabular}{ccrcr}
\hline \hline
ID  & Frequency & Period~~ & Amplitude  & S/N      \\
 & (1/d)& d~~~~~~ & (mmag) &   \\
\hline               
$f_{sector}$   & 0.035  &28.6$\pm$1.6 &  0.56 & 32.6      \\
$f_{orbit}$    & 0.078  &12.8$\pm$0.7 &  0.68 & 40.5   \\
$f_x$          & 0.136  & 7.4$\pm$0.4 &  0.56 & 31.3    \\
$f_0$          & 0.175  & 5.7$\pm$0.5 &  0.29 & 14.3   \\
$f_0+f_x$      & 0.306  & 3.3$\pm$0.2 &  0.18 & 8.2      \\
$2f_0$         & 0.341  & 2.9$\pm$0.1 &  0.20 & 6.9   \\
$f_1$          & 0.265  & 3.8$\pm$0.2 &  0.16 & 7.8      \\
$2f_1$         & 0.515  & 1.9$\pm$0.1 &  0.12 & 3.9    \\
\hline      
\end{tabular}
\end{table}

\begin{table*}
\centering
\caption{Adopted effective temperatures and luminosities for the stars discussed in Sect.~\ref{sect:discussion}.}
\label{tab:lum}
\begin{tabular}{lcrl}
\hline \hline
Star  & $T_{eff} (K)$ & $L/L_{\odot}$ & Source \\
\hline
HD 344787            &    5750$\pm$150  &   1435$\pm$147    &  This Work     \\       
Polaris              &    6000$\pm$100  &   2540$\pm$340    &  This Work     \\ 
BY Cas               &    6000$\pm$265  &   1700$\pm$41     &  \citet{Groenewegen2020}$^a$     \\
SZ Cas               &    5250$\pm$290  &   4607$\pm$1100   &  \citet{Groenewegen2020}$^a$  \\ 
RU Sct               &    5000$\pm$221  &   8367$\pm$2844   &  \citet{Groenewegen2020}$^a$ \\ 
ASAS 075842-2536.1   &    6300$\pm$100  &    126$\pm$58     &  \citet{catanzaro20}     \\
ASAS 131714-6605.0   &    6300$\pm$100  &    640$\pm$170    &  \citet{catanzaro20}     \\
V1033 Cyg            &    5860$\pm$100  &    742$\pm$192    &  \citet{catanzaro20}     \\
V371 Per             &    6000$\pm$100  &    949$\pm$668    &  \citet{catanzaro20}     \\
V363 Cas             &    6680$\pm$110  &    257$\pm$48     &  \citet{catanzaro20}     \\
\hline
\end{tabular}
\tablefoot{~a = the uncertainty in luminosity was increased with respect to the original value to include the error on the {\it Gaia} DR2 parallax.}\end{table*}

\section{HD\,344787 versus Polaris}\label{sect:comparison}

The two DCEPs addressed in our paper can be compared from different points of view and perspectives.

\subsection{Position in the Hertzsprung-Russell diagram (HRD).} 

The $T_{\rm eff}$ and luminosity values  derived in Sect.\ref{sect:dataAnalysis} were used to locate HD\,344787 and Polaris in the HRD, as shown in Fig.~\ref{fig:hr}. The figure includes the ISs for F and FO modes. 
DCEPs with metal abundances   Z=0.02 \citep{Desomma2020} and Z=0.008 \citep{Bono2001}, respectively. These values bracket the metallicity of the Sun used here, Z=0.0152, according to the PARSEC evolutionary tracks by \citet{Bressan2012}, which are  also displayed in the figure (for Z=0.014,  Y=0.273) as a reference for the typical masses expected from stellar evolution. The position in the HRD of the two stars is close but because of the small errors, significantly different in luminosity and $T_{\rm eff}$. According to \citet{Anderson2018}, Polaris is a 7 $M_{\odot}$ star at the hot boundary of the first-crossing IS. This seems confirmed by the location of the star in the HRD (see Fig.~\ref{fig:hr}), very close to the first-overtone blue edge (FOBE) of the Z=0.008 IS predicted by pulsation models, 
with a mass of about 6 $M_{\odot}$. Our  lower mass estimate 
is mainly due to the different parallax value  adopted here \citep[7.54$\pm$0.11 mas,][]{vanLeeuwen2007} compared to  \citet{Anderson2018}, who  adopted the parallax of Polaris B measured with the  HST \citep[6.26$\pm$0.24 mas,][]{Bond2018}.  
Our choice is also agrees very well with the parallax of Polaris B as published by {\it Gaia} DR2, $\varpi=7.292\pm0.028$ mas and  EDR3 \citep[early data release 3][]{Gaia2016,Gaia2020}, $\varpi=7.2869\pm0.0178$ mas. On this basis, the position of Polaris in the HRD appears to be rather accurate and is not compatible with the very low dynamical mass 3.5$\pm$0.8 $M_{\odot}$ measured by \citet{Evans2018} through the analysis of the Polaris system orbit \citep[see also their detailed discussion of the mass of Polaris in comparison with that of the binary DCEP V1334 Cyg, which was accurately measured by][]{Gallenne2018}. However, the same authors stated that more data are needed to better define the orbit of Polaris. This means that its dynamical mass estimate might be consistent with the evolutionary estimate.

The position of HD\,344787 appears to be  redder and fainter than Polaris, indicating a slightly lower mass, on the order of 5.3-5.5 $M_{\odot}$. In more detail, HD 344787 is placed slightly off the first-overtone red edge (FORE) for  Z=0.008, and just above the  intersection with the fundamental blue edge (FBE) for Z=0.02. This location agrees with the fact that the stellar pulsation become fainter and that HD\,344787 is a multi-mode (F/FO) pulsator. Therefore the difference in $T_{\rm eff}$ between the two DCEPs (albeit significant only at $\sim$1$\sigma$ level) allows us to estimate the width of the FO IS, that is,  $\Delta T_{\rm eff} \sim 250$ K \citep[in good agreement with theoretical predictions by][]{Bono2001}.

\subsection{Pulsation properties.} 

The pulsation mode of Polaris has long  been debated in the literature. However, \citet{Anderson2018}  convincingly showed that the FO pulsation mode should be preferred over the F mode \citep[as proposed by e.g.][]{Turner2013} or the second-overtone (2O) mode \citep[see][]{Bond2018}. 
As pointed out in Sect.~\ref{sect:intro},  
the period of Polaris is increasing at a high rate, and even after the so-called "glitch"   occurred in about 1965 \citep[][]{Turner2009,Turner2013}, when the rate of increase slightly decreased, its positive increment remains compatible with a redward evolving star in the HRD that crosses the IS for the first time. The light pulsation amplitude of Polaris prior to the 1965 glitch showed a slow but steady abatement, with an abrupt decline after this event, reaching  a minimum amplitude in $V$ band of $\sim$0.025 mag around 1988 \citep[][]{Turner2009}. It is currently slightly increasing \citep[][]{Bruntt2008,Turner2009,Turner2013}, but might vanish completely in several thousand years if the pre-glitch trend continues \citep[][]{Turner2009}. The current radial velocity amplitudes have not  returned to the levels measured before the glitch \citep{Usenko2018}. The glitch corresponds to a sudden drop in pulsation period, and according to \citet{Turner2009}, it might be explained by the quick acquisition of about seven Jovian masses by Polaris. 
This fascinating scenario is not so far fetched: Observations of the past two decades from the ground and space have revealed that extra-solar planets are very common. Interestingly, the possible assimilation of planetary companion(s) for an intermediate-mass star such as Polaris is expected to occur during the first crossing of the IS, that is, the evolutionary phase when the star expands its envelope for the first time to become a red super-giant.

For HD\,344787, the TESS data confirm excitation of two frequencies in the star, which  \citet{Turner2010} have identified as F and FO pulsation modes. Similarly to Polaris, the star shows a very fast period increase and diminishing amplitude, which is barely detectable and has been recorded only because of the high precision of TESS data.   

This complex observational scenario shows that Polaris and HD\,344787 have indeed close similarities but also remarkable differences. 
Both stars have quickly increasing periods, which can naturally be explained by them being  at the first crossing of the IS and evolving redward in the HRD. 
The diminishing pulsation amplitudes are more striking in HD\,344787, whose pulsation is only still detectable through TESS precise photometry, whereas the amplitude of Polaris, as remarked earlier, has a complex behaviour and in addition to a secular abatement, is now slightly increasing. 
An alternative explanation of these observations comes from the suggestion that in general, overtone pulsators might have unusually strong period changes \citep[][]{Szabados1983}. Adding to this the discontinuous amplitude variation, we can speculate about a phenomenon that is related to pulsation instead of evolution alone. A possible connection of pulsation and period or amplitude change might be a high rate of pulsation-induced mass loss. However, as noted by \citet{Evans2018}, we currently have no significant evidence of mass loss driven by pulsation in Polaris, and we have no data for HD 344787. The question whether the rapid period changes in Polaris and HD 344787 are just 
an extreme example of overtone pulsation instead of rapid evolution in the  
first instability-strip crossing is important and requires further dedicated observations.

If the first crossing of the IS explains the similarities between Polaris and HD\,344787, their different positions in the HRD accounts for the differences. 
Polaris is still rather close to the FOBE and evolves 
towards the FORE in a region of the IS in which a very narrow FO zone is expected according to pulsation models, while HD\,344787 is likely between the FBE and the FORE in a region of the IS in which multi-mode pulsation is expected.  The fast redward evolution in the HRD in the case of HD\,344787 seems to lead the star to  not only exit the FO IS, but also the F mode IS and to quit pulsation, as testified by the almost undetectable pulsation, thus suggesting that the FRE is located very close to the FORE. This occurrence is not confirmed by the theoretical boundaries plotted in Fig.~\ref{fig:hr}, where HD\,344787 seems significantly bluer than the predicted FRE for the two chemical compositions we considered. However, as suggested by a number of theoretical investigations devoted to the effect of super-adiabatic convection at the position of the DCEP IS boundaries \citep[see e.g.][and references therein]{Fiorentino2007,Desomma2020}, the assumed convective efficiency that should reasonably be increased in the red part of the IS significantly affect the position of the FRE. In particular, this extreme boundary becomes bluer by about 300 K as the mixing length parameter adopted in the pulsation code increases from 1.5 to 1.8 or 1.9. This blueward shift would be enough to place HD\,344787 very close to the predicted FRE.\\


\subsection{Chemical composition.} 

The chemical patterns of the two stars are quite similar. Within the uncertainties, all the species show similar  content. The abundances are also very consistent with solar standard values, at least within the experimental errors. In particular, carbon and oxygen have normal abundances, and this could be interpreted as a further sign of first-crossing DCEPs. 

The criterion usually adopted for spectroscopic identification of such objects is to search for stars whose chemical composition does not reflect 
changes induced by the first dredge-up (1DU, hereafter), which  occurs later during the post-main-sequence evolution of intermediate-mass stars. 
When 1DU takes place, it 
brings incomplete CNO-cycle processed material from the inner layers to the stellar surface, modifying the initial atmospheric abundances of the CNO elements. In particular, carbon becomes deficient with respect to its initial abundance [C/H]\,=\,$-$0.3, nitrogen becomes overabundant [N/H]\,=\,0.3, and oxygen is expected to remain practically unchanged \citep[][and references therein]{luck01}. 

Unfortunately, we do not have the typical nitrogen lines used in DCEPs (i.e. \ion{N}{I} $\lambda \lambda$\,7463.3 \AA{} and 8629.16 \AA{}) within our spectral range, 
therefore we can only base our conclusions on carbon and oxygen. 
We did not find any peculiar signature in our spectra because carbon and oxygen departures from solar values are negligible. We conclude in favour of a first-crossing phase for both HD\,344787 and Polaris.

It is worth noting that the evolutionary phase of Polaris is debated and controversial in the literature. \citet{usenko05} concluded based on their abundance analysis that Polaris probably is at its third crossing of the IS. A similar conclusion was reported by \citet{Evans2018}, who stated that the Polaris abundances are compatible with a post-first$^{}$-DUP scenario. In contrast, based on the abundances, \citet{Anderson2018} was in favor of our conclusion that Polaris reaches the IS for the first time.
\section{Discussion}\label{sect:discussion}

In the previous section we compared 
Polaris and HD 344787. Now, we compare their properties with the characteristics of other DCEPs that are thought to be crossing the IS for the first time. 
We can divide them into two groups with different characteristics. 
The first group is composed of objects with a significant positive increase in period, and in addition to Polaris and HD\,344787, it includes the FO mode DCEP BY Cas\footnote{The other DCEP traditionally considered to be a first-crossing candidate 
together with BY Cas is DX Gem. However, \citet{Berdnikov2019b} recently showed that the period of DX Gem decreases, which means that the star is now considered  a second-crossing candidate.} \citep[see e.g. ][and references therein]{Turner2010,Berdnikov2019a} and the two F mode DCEPs SZ Cas and RU Sct \citep[see e.g. ][and references therein]{Usenko2015}. 

The second group is composed of DCEPs whose spectra show the \ion{Li}{I} 6707.766 \AA{} line. This feature is associated with a first crossing of the IS because lithium is expected to be destroyed by proton-capture after the 1DU, that is, at the beginning of the red giant branch (RGB) phase \citep{Iben1967}. Five stars with these characteristics are known so far in our Galaxy: ASAS J075842-2536.1, ASAS J131714-6605.0, V371 Per, V1033 Cyg, and V363 Cas  \citep{luck11,Kovtyukh2016,Kovtyukh2019,catanzaro20}. 


DCEPs belonging to the two groups are also displayed in the HRD in Fig.~\ref{fig:hr} using the effective temperatures and luminosities listed in Table~\ref{tab:lum}. An inspection of Fig.~\ref{fig:hr} reveals that of the DCEPs that are candidates for first crossing, those showing Li lines in their spectra appear to be 
significantly less luminous.  
If all DCEPs shown in 
Fig.~\ref{fig:hr} 
are indeed first crossing the IS\footnote{The presence of Li in  ASAS 131714-6605.0 can be explained in a different way, as this star does indeed  show CNO abundances as it had  already gone through the 1DU.}, the underluminosity of the Li-rich DCEPs might suggest 
that mass drives 
the presence or absence of Li in DCEP spectra. 
On the other hand,  
1DU is not the only mechanism that is able to deplete lithium because rotational mixing can deplete the Li abundance to 1\% in a fraction of the main-sequence (MS) lifetime if the star rotates fast enough during the MS phase \citep[e.g.][]{Brott2011}. 

On this basis, even though the statistics is still poor, we might  tentatively explain the difference between   Li-rich and Li-depleted candidate first-crossing DCEPs in terms of mass and rotational velocity of the MS progenitors: the first are generally less massive and rotate more slowly than the latter, which include the two stars discussed in this paper, HD\,344787 and Polaris.


\section{Conclusions}\label{sect:conclusions}

We investigated HD344787, a DCEP defined by \citet{Turner2010} as "a Polaris analogue that is even more interesting than Polaris". 
We analysed proprietary HARPS-N at the TNG high-resolution spectra of HD 344787 together with  archival spectra of Polaris obtained with the same instrumental setup. 
We found that the chemical pattern  of HD\,344787 is compatible with the solar pattern, and this also holds for  the 
carbon and oxygen abundances. 
This is a first indication that the star has not undergone 1DU and therefore likely is in its first passage of the IS. Very similar results were found from the analysis of Polaris, even though we note that the evolutionary status, age, and mass of this important star are still a matter of debate in the recent literature.

\begin{figure*}
\includegraphics[width=\textwidth]{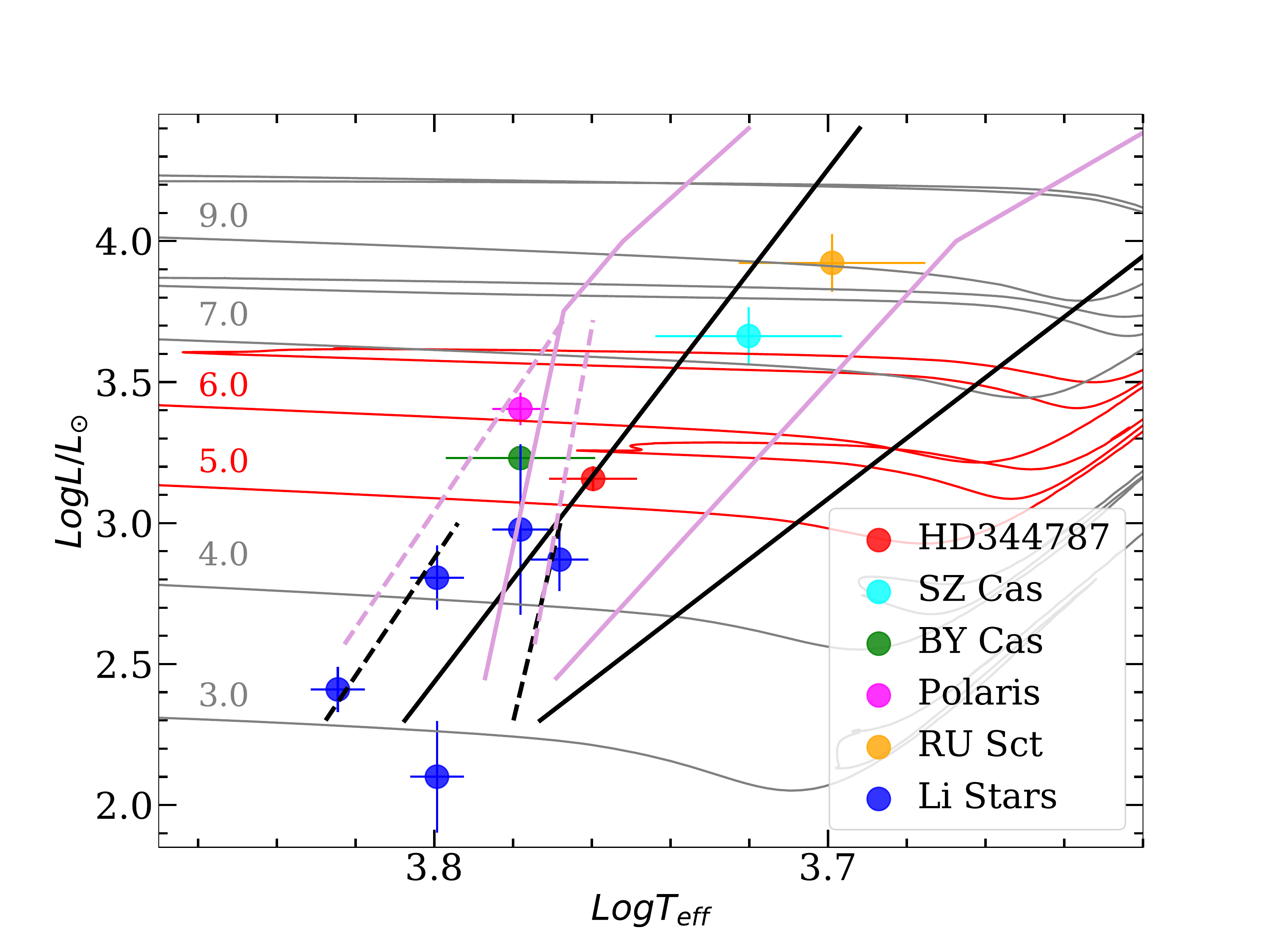}
\caption{HR diagram of HD\,344787 and Polaris together with the other candidate first-crossing DCEPs (see labels). The figure also reports MW Li-rich DCEPs from \citet{catanzaro20}. 
The ISs for F (solid lines) and FO (dashed lines) DCEPs are over-plotted in black for Z=0.02 \citep{Desomma2020} and in light violet for Z=0.008 \citep{Bono2001}. The evolutionary tracks by \citet{Bressan2012} with Z=0.014 and Y=0.273 for 5 and 6 $M_{\odot}$, encompassing the location of Polaris and HD 344787 in the HRD, are displayed in red. Additional tracks for 3, 4, 7, and 9 $M_{\odot}$ are over-plotted on the data in grey.}
\label{fig:hr}
\end{figure*}

As for the pulsation properties of HD\,344787, an analysis of the light curve observed by TESS  
has allowed us to confirm the prediction of \citet{Turner2010} that the very fast period increase together with the significant light amplitude decrease both point towards a vanishing pulsation in HD\,344787. The frequency analysis of the TESS light curve shows that the pulsation of the F and FO modes is at the level of only a few dozen mmag, which can only be measured from space. 
Consistently, no significant variation in radial velocity is measured in our spectra within the uncertainties ($\sim$100 m/s). 
The observational evidence we have gathered on HD\,344787 can be explained if the star is on the verge of crossing the FORE boundary and almost simultaneously the FRE, where convection is expected to damp  pulsation. This implies that the predicted FRE location displayed in Fig.~\ref{fig:hr} should be revised, increasing the efficiency of super-adiabatic convection in this red part of the IS, in agreement with previous suggestions \citep[see e.g.][and references therein]{Fiorentino2007,Desomma2020}. In this respect, HD\,344787 differs from Polaris, which is bluer by some 250 K and more likely close to the FOBE while crossing the IS for the first time. 

A comparison with other DCEPs that are thought   to be  on their first crossing of the IS suggests that these objects can be divided into two groups, depending on whether they show lithium in their atmospheres. This separation reflects a different position of each group  in the HRD, with lithium-rich DCEPs being systematically fainter than lithium-depleted DCEPs. Because lithium can be destroyed in the MS by rotational mixing, we speculate that the difference between the two groups indicates different MS progenitors, where lithium-rich DCEPs are the descendants of less massive, slowly rotating B stars in the MS. These results must be corroborated by more statistically significant  samples that will become available in the near future from spectroscopic surveys that will be carried out with forthcoming high-resolution, multi-object spectrographs such as WEAVE at the WHT \citep{Dalton2012}, 4MOST at VISTA \citep{deJong2012}, and MOONS at the VLT \citep{Cirasuolo2014}.


\begin{acknowledgements}

We wish to thank the anonymous Referee for their suggestions, which helped us to improve the paper.
We wish to thank the whole staff of the TNG for their strenuous efforts in carrying out the observations even in this difficult pandemic time. 
This work has made use of data from the European Space Agency (ESA) mission
{\it Gaia} (\url{https://www.cosmos.esa.int/gaia}), processed by the {\it Gaia} Data Processing and Analysis Consortium (DPAC,
\url{https://www.cosmos.esa.int/web/gaia/dpac/consortium}). Funding for the DPAC has been provided by national institutions, in particular the institutions participating in the {\it Gaia} Multilateral Agreement.
In particular, the Italian participation
in DPAC has been supported by Istituto Nazionale di Astrofisica
(INAF) and the Agenzia Spaziale Italiana (ASI) through grants I/037/08/0,I/058/10/0, 2014-025-R.0, and 2014-025-R.1.2015 to INAF (PI M.G. Lattanzi). This paper includes data collected by the TESS mission. Funding for the TESS mission is provided by NASA’s Science Mission directorate. The research leading to these results has been supported by the Lend\"ulet LP2014-17 and LP2018-7/2020 grants of the Hungarian Academy of Sciences and by the grant KH\_18 130405 of the 
Hungarian National Research, Development and Innovation Office (NKFIH).
L.M. was supported by the Premium Postdoctoral Research Program of the Hungarian Academy of Sciences.
This research made use of  NASA's Astrophysics Data System. 

\end{acknowledgements}

%
%

%


\end{document}